\documentclass[graybox]{svmult}

\usepackage{type1cm}        
\usepackage{makeidx}         
\usepackage{graphicx}        
\usepackage{multicol}        
\usepackage[bottom]{footmisc}
\usepackage{tikz}
\usetikzlibrary{arrows.meta, positioning, decorations.pathreplacing}

\usepackage{newtxtext}       %
\usepackage[varvw]{newtxmath}       
\usepackage{booktabs} 

\makeindex             

\usepackage[T1]{fontenc}
\usepackage{graphicx}

\usepackage{amssymb}
\usepackage{dcolumn}
\usepackage{bm, color}
\usepackage{graphicx}
\usepackage{diagbox} 
\usepackage{subfig}
\usepackage{caption}
\usepackage{algpseudocode}
\usepackage{algorithmicx}
\usepackage{algorithm}
\usepackage{makecell}
\usepackage{multirow}
\usepackage{hyperref}
\usepackage{smartdiagram}
\usepackage{verbatim}
\hypersetup{
    colorlinks=true,
    linkcolor=blue,
    filecolor=blue,      
    urlcolor=blue,
    citecolor=blue,
}

\usepackage{quantikz}

\usepackage{tikz}

\usetikzlibrary{positioning,patterns}
\usetikzlibrary{decorations.pathreplacing,decorations.pathmorphing}
\colorlet{colred}{red!22}
\colorlet{colcyan}{cyan!22}
\colorlet{colV}{blue!40}
\colorlet{colBorder}{gray!70}
\tikzset
  {mybox/.style=
    {rectangle,rounded corners,drop shadow,minimum height=1cm,
     minimum width=2cm,align=center,fill=#1,draw=colBorder,line width=1pt
    },
   myarrow/.style=
    {draw=#1,line width=3pt,-stealth,rounded corners
    },
   mylabel/.style={text=#1}
  }
\newcommand{\True}{\textbf{true}}
\newcommand{\False}{\textbf{false}}

\renewcommand{\epsilon}{\varepsilon}

\renewcommand{\c}{\ensuremath{\mathcal{C}}}

\newcommand{\e}{\ensuremath{\mathcal{E}}}

\newcommand{\h}{\ensuremath{\mathcal{H}}}

 \newcommand{\tr}{{\rm Tr}} 
 \renewcommand{\a}{\ensuremath{\mathcal{A}}}

\newcommand{\at}[1]{\textcolor{teal}{AT: #1 :AT}}

\usepackage{tabularray}

\begin{document}
\title{Verifying Adversarial Robustness in Quantum Machine Learning: from theory to physical validation via a software tool}
\titlerunning{Verifying Adversarial Robustness in Quantum Machine Learning}
\author{Ji Guan\orcidID{0000-0002-3490-0029} and\\ Mingsheng Ying\orcidID{0000-0003-4847-702X}}
\institute{Ji Guan \at Key Laboratory of System Software (Chinese Academy of Sciences), Institute of Software, Chinese Academy of Sciences, Beijing 100190, China \email{guanj@ios.ac.cn}
\and Mingsheng Ying \at Centre for Quantum Software and Information, University of Technology Sydney, Ultimo NSW 2007, Australia \email{Mingsheng.Ying@uts.edu.au}}
%
\maketitle
\abstract{
As with classical neural networks, quantum machine learning (QML) models are vulnerable to small input perturbations that can significantly alter output predictions. Certifying the robustness of QML models, particularly on NISQ hardware, is therefore a fundamental step toward trustworthy quantum AI.
This chapter reviews our recently developed comprehensive formal framework for \textit{verifying adversarial robustness} in QML. {The core of this framework} is a fidelity-based robustness lower bound computable directly from the measurement outcome distribution, which enables both formal verification and empirical estimation on real quantum devices. Additionally, the optimal bound can be computed via semidefinite programming (SDP) with full knowledge of the quantum machine learning models.  We incorporate these results into: (1) an efficient formal verification framework; (2) \textsc{VeriQR}, the first dedicated QML robustness verification tool; and (3) the first experimental benchmark of quantum adversarial robustness on a 20-qubit superconducting processor. Together, these systematic advances enable scalable, physically grounded robustness evaluation of QML models.}

\section{Introduction}

Quantum machine learning (QML) has emerged as a promising paradigm at the intersection of quantum computing and artificial intelligence, offering the potential to accelerate data-driven tasks such as classification, pattern recognition, and quantum system identification~\cite{biamonte2017quantum,cerezo2022challenges}. Enabled by rapid progress in quantum hardware platforms—particularly superconducting circuits—the implementation and empirical study of QML algorithms is advancing swiftly~\cite{s41586-019-0980-2,Herrmann2022,Gong2023,Tacchino2019,Huang2021,huang2021quantum,zhang2024quantum,chen2025quantum}. Industrial momentum is also building; for instance, Google has introduced \textit{TensorFlow Quantum}, integrating quantum circuit training into the mainstream classical ML framework TensorFlow~\cite{broughton2020tensorflow}.

However, like their classical counterparts, QML models are susceptible to \emph{adversarial perturbations}—subtle modifications to the input quantum state that can cause incorrect predictions or degrade performance~\cite{Lu2020,Liu2020}. These perturbed inputs, known as \emph{adversarial examples}, present a serious threat to the reliability and security of quantum AI, particularly in the noisy intermediate-scale quantum (NISQ) era where hardware noise is pervasive. Ensuring that QML models are robust against such perturbations is therefore a pressing concern~\cite{franco2024predominant}. 

While adversarial robustness has been widely studied in classical deep learning~\cite{huang2011adversarial,goodfellow2014explaining}, its quantum counterpart poses distinct challenges~\cite{nielsen2010quantum}. Quantum states are probabilistic, subject to decoherence and readout noise, and constrained by physical laws that restrict allowable perturbations to those preserving trace and positivity. This necessitates new verification frameworks grounded in quantum-specific geometry, such as fidelity-based distances.

In this chapter, we review our recently developed comprehensive framework for \textbf{verifying adversarial robustness in quantum machine learning}, grounded in rigorous theoretical analysis and experimental validation supported by a software tool. Our approach spans three interlinked dimensions:

\begin{itemize}
    \item \textbf{Theoretical Foundations.} We formalize robustness using fidelity-based metrics and introduce a sound, instance-specific \emph{robustness lower bound} that can be computed directly from the measurement outcome distribution~\cite{guan2021robustness}. This bound enables certification even in black-box scenarios without model access, which is used in the physical validation. Additionally, we formulate the exact robustness radius as a semidefinite program (SDP), enabling provably optimal verification when the model is known. Based on these results, we design a set of verification algorithms that certify robustness under various information assumptions.
    
    \item \textbf{Algorithmic Realization.} We implement these ideas in \textsc{VeriQR}~\cite{lin2024obustness}, the first dedicated tool for QML robustness verification. \textsc{VeriQR} supports both exact (sound and complete) and approximate robustness verification algorithms, and can identify quantum adversarial examples that may be used for adversarial training. The tool accepts models in OpenQASM 2.0 format and simulates realistic noise to reflect NISQ hardware characteristics, offering a unified benchmarking framework compatible with multiple quantum platforms.

    \item \textbf{Experimental Validation.} To assess practical viability, {the first physical benchmark of QML robustness was conducted on a 20-qubit superconducting processor~\cite{zhang2025experimental} based on our theoretical framework and using the \textsc{VeriQR} tool. It was demonstrated that the fidelity-based lower bound provides a tight and stable estimate of robustness in the presence of real quantum noise and validates its correlation with robustness upper bounds obtained by adversarial attack methods and SDP-derived optimal bounds.}
\end{itemize}

Together, the above components establish a scalable, hardware-compatible, and formally grounded approach to robustness in quantum machine learning. They show that rigorous certification is feasible on near-term devices and can be integrated seamlessly into the QML development lifecycle.

\textbf{Organization.} The remainder of this review is organized as follows. Section~\ref{sec:prelim} introduces the necessary preliminaries, including quantum classifiers, fidelity-based distance measures, and the formal definition of the robustness verification problem. Section~\ref{sec:formalism} develops a series of robustness bounds that form the theoretical foundation for solving the verification task. Section~\ref{Sec:algorithms} presents verification algorithms derived from these bounds. Section~\ref{sec:veriqr} describes the design and functionality of the \textsc{VeriQR} tool, which implements these verification techniques. Section~\ref{sec:benchmark} details the experimental setup and results from benchmarking adversarial robustness on real superconducting quantum hardware using the robustness bounds. Finally, Section~\ref{sec:outlook} concludes with a discussion of open problems and future research directions.

\section{Preliminaries}
\label{sec:prelim}

This section presents the foundational concepts essential for verifying adversarial robustness in quantum machine learning. We begin by introducing the structure and semantics of quantum classifiers, followed by the definition of fidelity-based distance measures used to quantify adversarial perturbations. These perturbations can lead to adversarial examples, which are input states that induce misclassification. Building on these definitions, we then formalize the robustness verification problem, which forms the central focus of this chapter.

\subsection{Quantum Classifiers}
\begin{figure}[ht]
\centering
\begin{tikzpicture}[
  node distance=2.5cm and 1.8cm,
  every node/.style={font=\normalsize},
  process/.style={draw, thick, rounded corners, minimum height=1cm, minimum width=1.8cm, align=center},
  arrow/.style={-{Latex[length=3mm]}, thick}
  ]

\node (rho) {\(\rho \in \mathcal{D}(\mathcal{H})\)};
\node[process, right=of rho, fill=blue!15] (channel) {\(\mathcal{E}\)};
\node[process, right=of channel, fill=green!20] (povm) {\(\{M_c\}_{c\in\c}\)};
\node[right=of povm] (output) {\(c = \a(\rho)\)};

\draw[arrow] (rho) -- (channel);
\draw[arrow] (channel) -- (povm);
\draw[arrow] (povm) -- (output);

\end{tikzpicture}
\caption{\textbf{Quantum classifier pipeline.} The input quantum state \(\rho\) is processed by a quantum channel \(\mathcal{E}\), followed by measurement via a POVM \(\{M_c\}_{c\in\c}\), to produce a classical class label \(c = \a(\rho)\).}
\label{fig:classifier_pipeline}
\end{figure}
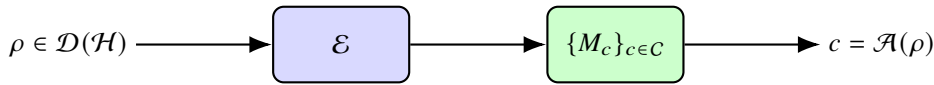
Let $\mathcal{H}$ be a $2^n$-dimensional Hilbert space on an $n$-qubit quantum system. A \emph{quantum state} $\rho \in \mathcal{D}(\mathcal{H})$ is a positive semidefinite operator ($\rho\succeq 0$) on $\h$ with trace one ($\tr(\rho)=1$). Here $\mathcal{D}(\mathcal{H})$ represents the set of quantum states on $\h$.  

As illustrated in Fig.~\ref{fig:classifier_pipeline}, a \emph{quantum classifier} is a quantum algorithm that takes quantum input states and produces classical output labels, corresponding to predefined classes of interest. Formally, a quantum classifier over the Hilbert space $\mathcal{H}$ is defined as a pair:
\[
\a = (\mathcal{E}, \{M_c\}_{c \in \mathcal{C}}),
\]
where:
\begin{itemize}
\item $\mathcal{E}: \mathcal{D}(\mathcal{H}) \to \mathcal{D}(\mathcal{H})$ is a \emph{quantum channel}, a completely positive and trace-preserving (CPTP) map, that models the quantum evolution (including noise) of the algorithm;
\item $\{M_c\}{c \in \mathcal{C}}$ is a \emph{positive operator-valued measure} (POVM), where each $M_c \succeq 0$ and $\sum_c M_c = I$, used to perform quantum measurement and extract classical outputs;
\item $\mathcal{C}$ is a finite set of possible output labels representing the classification categories.
\end{itemize}

Given an input quantum state $\rho \in \mathcal{D}(\mathcal{H})$, the classifier outputs a label determined by the most probable measurement outcome:
\[
\a(\rho) := \arg\max_{c \in \mathcal{C}} \mathrm{Tr}[M_c \mathcal{E}(\rho)],
\]
where $\mathrm{Tr}[M_c \mathcal{E}(\rho)]$ is the probability of obtaining outcome $c$ upon measuring the output state $\mathcal{E}(\rho)$ of $\e$ with the POVM $\{M_c\}_{c \in \mathcal{C}}$.

\subsection{Distance Between Quantum States}

A central component of robustness analysis is the choice of a metric to quantify the similarity between quantum states. In the context of quantum computing, the standard metric used to characterize adversarial perturbations is based on \emph{quantum fidelity}~\cite{nielsen2010quantum}.

\begin{definition}[Fidelity]~\cite{nielsen2010quantum}  
Let \(\rho, \sigma \in \mathcal{D}(\mathcal{H})\) be two quantum states. The fidelity between \(\rho\) and \(\sigma\) is defined as
\[
F(\rho, \sigma) := \left( \mathrm{Tr} \sqrt{\sqrt{\rho} \sigma \sqrt{\rho}} \right)^2.
\]
\end{definition}

Fidelity measures the closeness of two quantum states. It ranges from 0 to 1, where \(F(\rho, \sigma) = 1\) if and only if \(\rho = \sigma\), and \(F(\rho, \sigma) = 0\) when \(\rho\) and \(\sigma\) have orthogonal support.

\begin{definition}[Fidelity Distance]  
The \emph{fidelity distance} (also called \emph{infidelity}) between two quantum states is defined as
\[
D_F(\rho, \sigma) := 1 - F(\rho, \sigma).
\]
\end{definition}

Fidelity distance provides a quantitative measure of how distinguishable two quantum states are under optimal measurements~\cite{nielsen2010quantum}. In robustness verification, it serves as a natural metric for evaluating the magnitude of adversarial perturbations: smaller distances imply higher similarity, while larger distances indicate more significant deviations from the original state.
\subsection{Adversarial Robustness}

To assess the adversarial robustness of a quantum classifier at a given input state \(\rho\), we require that all quantum states sufficiently close to \(\rho\) be classified identically to \(\rho\). A violation of this condition indicates the presence of an \emph{adversarial example}—a nearby state that is misclassified.

\begin{definition}[Adversarial Example]
Let \(\a\) be a quantum classifier, \(\rho \in \mathcal{D}(\mathcal{H})\) an input state, and \(\epsilon > 0\) a perturbation threshold. A quantum state \(\sigma\) is called an \emph{\(\epsilon\)-adversarial example} of \(\rho\) if
\[
\a(\sigma) \ne \a(\rho) \quad \text{and} \quad D_F(\rho, \sigma) \le \epsilon.
\]
If such a state \(\sigma\) exists, then \(\epsilon\) is referred to as an \emph{adversarial perturbation} of \(\rho\).
\end{definition}

Here, the condition \(\a(\sigma) \ne \a(\rho)\) indicates that \(\sigma\) is classified differently from \(\rho\), while the condition \(D_F(\rho, \sigma) \le \epsilon\) ensures that \(\sigma\) is sufficiently close to \(\rho\) in fidelity distance. Thus, an \(\epsilon\)-adversarial example is a small perturbation of \(\rho\) that causes the classifier to mispredict.

Note that this definition implicitly assumes that \(\rho\) is correctly classified. If \(\a(\rho)\) is already incorrect, identifying \(\sigma\) such that \(\a(\sigma) \ne \a(\rho)\) no longer signifies a robustness issue but a correctness one. Therefore, in the subsequent discussions, we restrict attention to correctly classified input states.

The absence of adversarial examples within a given neighborhood defines robustness:

\begin{definition}[Adversarial Robustness]\label{def:adversary}
A quantum classifier \(\a\) is said to be \emph{\(\epsilon\)-robust} at state \(\rho\) if there exists no \(\epsilon\)-adversarial example of \(\rho\).
\end{definition}

Equivalently, \(\a\) is \(\epsilon\)-robust at \(\rho\) if
\[
\forall \sigma \in \mathcal{D}(\mathcal{H}), \quad D_F(\rho, \sigma) \le \epsilon \ \Rightarrow \ \a(\sigma) = \a(\rho).
\]

This notion allows us to define the exact robustness radius of a quantum state:

\begin{definition}[Robustness Radius]
Let \(\a\) be a quantum classifier and \(\rho\) a correctly classified input state. The \emph{robustness radius} of \(\rho\), denoted \(\epsilon^*(\rho)\), is the maximum value \(\epsilon\) such that \(\a\) is \(\epsilon\)-robust at \(\rho\):
\[
\epsilon^*(\rho) := \sup_{\substack{\sigma \in \mathcal{D}(\mathcal{H}) \\ \a(\sigma) = \a(\rho)}} D_F(\rho, \sigma).
\]
\end{definition}

Intuitively, \(\epsilon^*(\rho)\) quantifies the largest allowable fidelity degradation that preserves the classifier’s decision. For any \(\epsilon > \epsilon^*(\rho)\), there exists an \(\epsilon\)-adversarial example \(\sigma\) that is misclassified relative to \(\rho\).

The robustness radius gives rise to the central verification task addressed in this chapter:

\begin{problem}[Robustness Verification Problem]
\label{prob:certification}
Given a quantum classifier \(\a\), an input state \(\rho \in \mathcal{D}(\mathcal{H})\), and a threshold \(\epsilon > 0\), determine whether
\[
\epsilon \leq \epsilon^*(\rho).
\]
If so, \(\a\) is \(\epsilon\)-robust at \(\rho\); otherwise, \(\epsilon\) is an adversarial perturbation, and a violating state \(\sigma\) can be returned as an \(\epsilon\)-adversarial example.
\end{problem}

The objective is to determine whether adversarial examples exist within an \(\epsilon\)-fidelity ball centered at \(\rho\). The remainder of this chapter develops exact and approximate approaches to solving this verification problem.

Finally, robustness can be aggregated across a dataset to evaluate a classifier's overall robustness:

\begin{definition}[Robust Accuracy]\label{def:robust_ac}
Let \(\a\) be a quantum classifier. The \emph{\(\epsilon\)-robust accuracy} of \(\a\) is the proportion of correctly classified input states in the dataset that are also \(\epsilon\)-robust.
\end{definition}

This definition extends the notion of robustness from individual quantum states to an entire dataset. Depending on the application, the dataset may consist of training samples, validation samples, or a combination of both, under the assumption that all included quantum states are correctly labeled.

\section{Robustness Bounds}
\label{sec:formalism}

This section presents a series of robustness bounds that address the robustness verification problem formulated in Problem~\ref{prob:certification}. These bounds characterize the robustness radius $\varepsilon^*(\rho)$ through exact, under-approximate, and over-approximate techniques. We derive these bounds using a variety of methods, including semidefinite programming, analysis of measurement outcome distributions, and adversarial attack construction.
\subsection{Optimal Robustness Bound via Semidefinite Programming}
We first show that the robustness radius $\varepsilon^*(\rho)$ can be exactly computed using \emph{semidefinite programming (SDP)}, assuming full knowledge of the quantum classifier.

Given the explicit description of the quantum channel $\mathcal{E}$ and the POVM $\{M_c\}_{c\in\mathcal{C}}$, the robustness radius can be formulated as an SDP~\cite{guan2021robustness}:

\begin{theorem}[Optimal Robustness Bound via SDP~\cite{guan2021robustness}]\label{Thm:optimal}
Let $\a= (\mathcal{E}, \{M_c\}_{c \in \mathcal{C}})$ be a quantum classifier. The exact robustness radius is given by
\[
\varepsilon^*(\rho) = \min_{\substack{c \in \mathcal{C} \\ c \ne \a(\rho)}} \epsilon^*_c(\rho),
\]
where each $\epsilon^*_c(\rho)$ is the solution to the following SDP:
\[
\begin{aligned}
\text{minimize:} \quad & D_F(\rho, \sigma) \\
\text{subject to:} \quad & \sigma \succeq 0, \\
& \mathrm{Tr}(\sigma) = 1, \\
& \mathrm{Tr}[(M_{\a(\rho)} - M_c)\mathcal{E}(\sigma)] \le 0.
\end{aligned}
\]
If this SDP is infeasible for some $c$, then $\epsilon^*_c(\rho) = \infty$, indicating that no adversarial example of $\rho$ exists which is misclassified as class $c$.
\end{theorem}

This SDP formulation allows exact and efficient computation of the robustness radius using convex optimization. The tractability of this approach is rooted in a fundamental principle of quantum mechanics: the linearity of quantum operations. Both the quantum channel $\mathcal{E}$ and the measurement process are linear maps, which leads to a convex feasible set and objective in the optimization problem.

This stands in contrast to classical neural networks, where nonlinearities such as ReLU activations or pooling layers make the robustness verification problem non-convex. As a result, certifying adversarial robustness in classical settings is NP-complete~\cite{katz2017reluplex}, and typically requires approximations or linear encodings of nonlinear functions (e.g., NSVerify~\cite{lomuscio2017approach}, MIPVerify~\cite{tjeng2017evaluating}, ILP~\cite{bastani2016measuring}, ImageStar~\cite{tran2020verification}).

However, computing $\varepsilon^*(\rho)$ using Theorem~\ref{Thm:optimal} assumes full access to the internal structure of the classifier, including its noise-free quantum evolution. In practice, especially on near-term noisy intermediate-scale quantum (NISQ) devices, this assumption rarely holds due to unknown or device-specific noise effects.

To address this, we introduce lower and upper bounds on $\varepsilon^*(\rho)$ that can be estimated under partial or empirical knowledge of the model. While the SDP method offers a gold-standard verification backend, these bounds provide practical alternatives for applications where exact model information is unavailable. In the following subsections, we present such bounds and discuss how they partially solve the robustness verification problem (Problem~\ref{prob:certification}).

\subsection{Robustness Lower Bound via Measurement Distribution}
We now present a certified lower bound for the robustness radius $\epsilon^*(\rho)$ based on the measurement outcome distribution of a quantum classifier.

Let $\a = (\mathcal{E}, \{M_c\}_{c \in \mathcal{C}})$ be a quantum classifier. For a given input state $\rho \in \mathcal{D}(\mathcal{H})$, denote the probability of observing class $c$ as 
\[
p_c^\rho := \mathrm{Tr}[M_c \mathcal{E}(\rho)].
\]
Let $c^* := \a(\rho)$ be the predicted class label of $\rho$. Then, the measurement distribution $\{p_c^\rho\}_{c \in \mathcal{C}}$ enables the following certified lower bound:

\begin{theorem}[Robustness Lower Bound from Measurement Distribution~{\cite{guan2021robustness}}]
\label{thm:lower}
Let $\rho \in \mathcal{D}(\mathcal{H})$ and $c^* = \a(\rho)$. Then
\[
\varepsilon_{\mathrm{RLB}}(\rho) := \min_{c \ne c^*} \frac{1}{2} \left( \sqrt{p_{c^*}^\rho} - \sqrt{p_c^\rho} \right)^2
\]
is a certified robustness lower bound: for all $\sigma$ such that $D_F(\rho, \sigma) \leq \varepsilon_{\mathrm{RLB}}(\rho)$, it holds that $\a(\sigma) = \a(\rho)$.
\end{theorem}

This robustness lower bound has the following key features:

\begin{itemize}
  \item \textbf{Efficient to Compute.} The bound depends only on the measurement probabilities $\{p_c^\rho\}_{c \in \mathcal{C}}$, and can be evaluated directly from measurement outcomes of $\rho$ without searching for adversarial perturbations. This efficiency allows for fast robustness certification and dataset-level evaluation of robust accuracy.
  
  \item \textbf{Model-agnostic:} Since it requires no access to the internal structure of $\mathcal{E}$, this bound is particularly suited for hardware-level evaluation. In real-device settings, one can estimate $p_c^\rho$ by repeated execution of $\mathcal{E}$ on quantum hardware and compute $\varepsilon_{\mathrm{RLB}}(\rho)$ from the empirical outcome distribution.

\end{itemize}

\subsection{Robustness Upper Bound via Attack Generation}
\label{sec:upper_bound}

In contrast to certified lower bounds, robustness upper bounds estimate the minimum perturbation required to fool a quantum classifier. These bounds can be obtained through adversarial attack strategies and provide empirical insights into the model’s worst-case vulnerability.

\begin{definition}[Empirical Robustness Upper Bound]
\label{def:rub}
Let $\rho \in \mathcal{D}(\mathcal{H})$ be an input quantum state. An \emph{adversarial attack method} constructs a perturbed state $\sigma_{\text{adv}}$ such that:
\[
\a(\sigma_{\text{adv}}) \ne \a(\rho), \quad \text{and} \quad \varepsilon_{\mathrm{RUB}}(\rho) := D_F(\rho, \sigma_{\text{adv}}),
\]
where $D_F$ is the fidelity distance. Then, $\varepsilon_{\mathrm{RUB}}(\rho)$ serves as an \emph{empirical robustness upper bound} for $\varepsilon^*(\rho)$.
\end{definition}

Since any adversarial example must lie outside the certified robust region, the inequality
\[
\varepsilon^*(\rho) < \varepsilon_{\mathrm{RUB}}(\rho)
\]
always holds. Thus, $\varepsilon_{\mathrm{RUB}}$ provides an over-approximation of the robustness radius. Although $\varepsilon_{\mathrm{RUB}}$ is not a certified bound, it provides practical evidence of vulnerability and complements robustness lower bounds. It is especially valuable in real-device settings where exact robustness certification via SDP is infeasible.

In classical adversarial learning, numerous algorithms have been developed to craft adversarial examples~\cite{ren2020adversarial}. These methods are typically categorized as either white-box or black-box attacks, depending on the attacker's access to model parameters. A widely used white-box approach is the \emph{Fast Gradient Sign Method (FGSM)}~\cite{Goodfellow2014}, which perturbs a legitimate input $\boldsymbol{x}$  as follows:
\begin{equation}\label{eq:FGSM}
\boldsymbol{x}' = \boldsymbol{x} + \epsilon \cdot \operatorname{sgn}({\nabla}_{\boldsymbol{x}} \mathcal{L}),
\end{equation}
where $\epsilon$ denotes the perturbation magnitude, ${\nabla}_{\boldsymbol{x}} \mathcal{L}$ is the gradient of the loss function $\mathcal{L}$ with respect to the input and the function $\operatorname{sgn}(\cdot)$ stands for the sign function to extract the direction of a gradient without its magnitude.

In the quantum setting, generating analogous adversarial examples requires gradient estimation using the parameter-shift rule~\cite{wierichs2022general}. This introduces significant computational overhead on quantum hardware: producing a single adversarial example typically demands $2N_s \cdot N_x$ circuit executions, where $N_s$ is the number of shots per measurement and $N_x = \dim(\boldsymbol{x})$ is the input dimension. Consequently, applying FGSM to high-dimensional quantum data becomes impractical, particularly in batch settings. This overhead highlights a major limitation of naive gradient-based attacks in quantum machine learning, especially on current noisy intermediate-scale quantum (NISQ) devices~\cite{Mitarai2018,schuld2019evaluating}.

To mitigate the overhead of full-gradient-based attacks, a localized variant of the FGSM, termed \emph{Mask FGSM}, has been proposed for efficient adversarial sample generation in quantum machine learning (QML) experiments~\cite{zhang2025experimental}. This method restricts perturbations to a strategically selected sparse subset of input features, specified by a binary mask $\mathcal{M} = (m_1, m_2, \ldots, m_{\operatorname{dim}(\boldsymbol{x})})^T$. The mask identifies which components of the input are perturbed, reducing both computational cost and the influence of noisy gradient estimates in experimental settings.

Given a legitimate input $\boldsymbol{x}$, adversarial examples are generated by applying a perturbation vector $\boldsymbol{\delta}$ such that $\boldsymbol{x}' = \boldsymbol{x} + \boldsymbol{\delta}$, where each component $\delta_i$ is computed as:
\begin{equation}\label{eq:mask_FGSM}
\delta_i = 
\begin{cases}
\epsilon \cdot \operatorname{sgn}\left(  \frac{\partial \mathcal{L} }{\partial x_i }  \right), & \text{if } m_i = 1, \\
0, & \text{if } m_i = 0,
\end{cases}
\end{equation}
with $\epsilon$ denoting the perturbation strength and $\mathcal{L}$ the loss function. The mask $\mathcal{M}$ is constructed by ranking gradient magnitudes, allowing targeted perturbation of the most influential input features (see~\cite{zhang2025experimental} for implementation details).

This sparse attack strategy achieves a favorable trade-off between efficiency and effectiveness, as confirmed in the quantum hardware experiments on EMNIST and LCEI classification tasks~\cite{zhang2025experimental}, which will be detailed in Section~\ref{sec:benchmark}.


\begin{figure}[ht]
\centering
\begin{tikzpicture}[scale=1.25]

\def\rLB{1.5}
\def\rStar{2.3}
\def\rUB{3}

\draw[fill=red!10, draw=red!60, thick] (0,0) circle (\rUB);
\node at (0,-\rStar-0.16) {\footnotesize Adversarial attack region  \(\varepsilon_{\mathrm{RUB}}\)};

\draw[fill=green!10, draw=green!60, thick] (0,0) circle (\rStar);
\node at (0,-\rLB-0.3) {\footnotesize Optimal robustness \(\varepsilon^*\)};

\draw[fill=blue!10, draw=blue!60, thick] (0,0) circle (\rLB);
\node at (0,-\rLB+0.6) {\footnotesize Certified region \(\varepsilon_{\mathrm{RLB}}\)};

\filldraw (0,0) circle (1pt) node[above left=-0.1pt] {\(\rho\)};

\draw[->, thick, blue!80!black] (0,0) -- (\rLB-0.3,0.9) node[midway, above=4pt] {\(\varepsilon_{\mathrm{RLB}}\)};
\draw[->, thick, green!80!black] (0,0) -- (\rStar,0) node[midway, above right=-2pt] {\(\varepsilon^*\)};
\draw[->, thick, red!80!black] (0,0) -- (\rUB-0.3,-1.3) node[midway, above right=-2pt] {\(\varepsilon_{\mathrm{RUB}}\)};

\end{tikzpicture}
\caption{Visualization of robustness bounds as nested fidelity-distance regions around the input state \(\rho\). The certified lower bound \(\varepsilon_{\mathrm{RLB}}\) defines a guaranteed-safe region, the upper bound \(\varepsilon_{\mathrm{RUB}}\) corresponds to successful adversarial attacks, and the optimal robustness \(\varepsilon^*\) lies in between.}
\label{fig:robustness_bounds_fidelity_scale}
\end{figure}
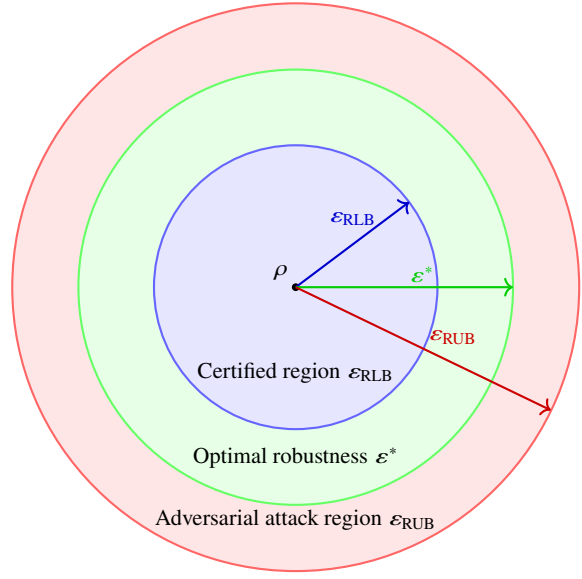

\subsection{Visualizing the Bounds}

Based on the certified lower bound and the empirical upper bound presented in the previous subsections, we can formalize their relationship through the following result:

\begin{theorem}[Sandwich Robustness Bound]
Given a quantum input state \(\rho\), a certified lower bound \(\varepsilon_{\mathrm{RLB}}(\rho)\) (Theorem~\ref{thm:lower}), and an adversarially generated state \(\sigma_{\text{adv}}\), we have:
\begin{equation}\label{Sand}
\varepsilon_{\mathrm{RLB}}(\rho) \le \varepsilon^*(\rho) \le \varepsilon_{\mathrm{RUB}}(\rho),
\end{equation}
where \(\varepsilon_{\mathrm{RUB}}(\rho) = D_F(\rho, \sigma_{\text{adv}})\).
\end{theorem}

This relation is illustrated in Figure~\ref{fig:robustness_bounds_fidelity_scale}. Each quantity in inequality~\eqref{Sand} has a distinct role in robustness analysis:
\begin{itemize}
  \item \(\varepsilon_{\mathrm{RLB}}(\rho)\): a certified lower bound used for formal robustness guarantees;
  \item \(\varepsilon^*(\rho)\): the exact robustness radius, computable via SDP;
  \item \(\varepsilon_{\mathrm{RUB}}(\rho)\): an empirical upper bound derived from adversarial attacks.
\end{itemize}

\textbf{Tightness Assessment.}  
The gap \(\Delta := \varepsilon_{\mathrm{RUB}}(\rho) - \varepsilon_{\mathrm{RLB}}(\rho)\) quantifies the precision of the robustness estimation. In the superconducting hardware experiments presented later~\cite{zhang2025experimental}, this certified lower bound is compared against an empirical robustness upper bound $\varepsilon_{\mathrm{RUB}}(\rho)$, obtained using gradient-based adversarial attacks (introduced in the next section). The observed gap between the two bounds is typically less than $3 \times 10^{-3}$, demonstrating that $\varepsilon_{\mathrm{RLB}}(\rho)$ provides a tight and practically useful certificate of robustness. 

\section{Robustness Verification Algorithms}\label{Sec:algorithms}


In this section, we present several algorithms for formally verifying the robustness of quantum classifiers, building upon the theoretical lower and optimal bounds introduced in the previous section.
\subsection{State Robustness Verification}
We begin by considering the robustness of a specific quantum input state \(\rho\) under a quantum classifier \(\a\). The robustness verification problem (Problem~\ref{prob:certification}) is to determine whether \(\a\) is \(\epsilon\)-robust at \(\rho\) for a given threshold \(\epsilon > 0\). 

Note that once the exact robustness radius \(\epsilon^*(\rho)\) is computed, verifying \(\epsilon\)-robustness reduces to a simple comparison: \(\a\) is \(\epsilon\)-robust at \(\rho\) if and only if \(\epsilon \leq \epsilon^*(\rho)\). This observation, together with Theorem~\ref{Thm:optimal}, leads to Algorithm~\ref{Algorithm}, which simultaneously determines the robustness status of \(\rho\) and computes the exact robustness radius \(\epsilon^*(\rho)\). 

\begin{algorithm}[htbp]
\caption{StateRobustnessVerifier($\a,\epsilon,\rho$)}
\label{Algorithm}
    \begin{algorithmic}[1]
    \Require $\a=(\e,\{M_{c}\}_{c\in\c})$ is a quantum classifier, $\epsilon < 1$ is a real number, $\rho$ is an input state of $\a$
    \Ensure \True{} indicates  $\a$ is $\epsilon$-robust at $\rho$ or \False{} with an adversarial example $\sigma$ indicates $\a$ is not $\epsilon$-robust at $\rho$   
    \ForAll{$c\in\c$ and $c\not=\a(\rho)$}\label{Algorithm:line:count}
    \State\label{Algorithm:line:SDP} By a SDP solver, compute $\epsilon^*_{c}(\rho)$ with an optimal state $\sigma_{c}$ in the SDP of Theorem~\ref{Thm:optimal}
    \EndFor
    \State Let $\epsilon^*(\rho)=\min_c\epsilon^*_{c}(\rho)$ and $c^*=\arg\min_c\epsilon^*(\rho)$
    \If{$\epsilon^*(\rho)>\epsilon$}
    \State \Return \True{}
    \Else
    \State \Return \False{} and $\sigma_{c^*}$
    \EndIf
    \end{algorithmic}  
\end{algorithm}

The primary computational cost of Algorithm~\ref{Algorithm} lies in solving semidefinite programs (SDPs), as shown in Line~\ref{Algorithm:line:SDP}. These SDP instances scale as \(O(N^{6.5})\) when solved using interior-point methods~\cite{zhang2018sparse}, where \(N\) is the dimension of the Hilbert space \(\mathcal{H}\). Since the algorithm requires solving an SDP for each incorrect class label (i.e., \( |\c| - 1 \) times), the overall computational complexity is given by the following:

\begin{theorem}
The worst-case complexity of Algorithm~\ref{Algorithm} is \(O(|\c| \cdot N^{6.5})\), where \(N\) is the dimension of the input state \(\rho\), and \(|\c|\) is the number of class labels.
\end{theorem}

\begin{algorithm}[htbp]
\caption{RobustnessVerifier($\a,\epsilon,T$)}
\label{algo:robustcheck}
    \begin{algorithmic}[1]
    \Require  $\a=(\e,\{M_c\}_{c\in\c})$ is a well-trained quantum classifier, $\epsilon < 1$ is a real number, $T=\{(\rho_i,l_{i})\}$ is a dataset indicating $\rho_i$ be correctly classified into class $l_i$.
    \Ensure The  robust accuracy $RA$ and a set $R=\{<\sigma_j, i_{j}>\}$, where for each $j$, $\sigma_j$ is an $\epsilon$-adversarial example of $\rho_{i_j}$; $R$ can be an empty set if $\a$ is $\epsilon$-robust at any state in $T$.
    \State $R=\emptyset$ be an empty set. \Comment{Recording adversarial examples and corresponding indexes of states in dataset $T$}
    \ForAll{$(\rho_i,l_i)\in T$}
    \State Let \( p_c= \mathrm{Tr}[M_c \mathcal{E}(\rho_i)] \) be the measurement probability of observing outcome \( c \) for input \( \rho \).
    \State Obtain the lower bound $\varepsilon_{\mathrm{RLB}}(\rho_i)= \min_{c \ne c^*} \frac{1}{2} \left( \sqrt{p_{c^*}} - \sqrt{p_c} \right)^2$\Comment{Applying the robust bound in Theorem~\ref{thm:lower}}
    \If {$\epsilon>\varepsilon_{\mathrm{RLB}}(\rho)$} 
    \If{StateRobustnessVerifier $(\a,\epsilon,\rho_i) ==$ \False{}} 
    \State $\sigma$ be the output state of StateRobustnessVerifier $(\a,\epsilon,\rho_i)$ 
    \State $R=R\cup \{(\sigma, i)\}$
    \EndIf
    \EndIf
    \EndFor
    \State \Return $RA=1-\frac{|R|}{|T|}$, $R$ \Comment{$|R|=0$ if $R$ is a empty set}
    \end{algorithmic}
\end{algorithm}
\subsection{Classifier Robustness Verification}
We now turn our attention to verifying the robustness of a quantum classifier \(\a\) as a whole. Algorithm~\ref{algo:robustcheck} is developed for this purpose by combining the exact robustness computation procedure (Algorithm~\ref{Algorithm}) with the certified lower bound technique from Theorem~\ref{thm:lower}.

One key advantage of formal robustness verification in classical machine learning is its ability to identify counterexamples, i.e., adversarial examples, for specific inputs (see, e.g.,~\cite{tran2020verification,elboher2020abstraction,fremont2020formal,kwiatkowska2019safety}). This advantage is preserved in the quantum setting through Algorithm~\ref{algo:robustcheck}, which enables both verification and counterexample generation. In particular, this facilitates a natural extension of adversarial training~\cite{madry2017towards} to quantum machine learning. Specifically, when a quantum state \(\rho\) fails the \(\epsilon\)-robustness check, the algorithm automatically identifies an adversarial example \(\sigma\). By augmenting the training dataset with the misclassified pair \((\sigma, l)\), where \(l\) is the correct label, the classifier \(\a\) can be retrained to enhance its robustness. This iterative refinement process allows quantum models to gradually become more resilient to adversarial perturbations.

To analyze the complexity of Algorithm~\ref{algo:robustcheck}, we begin by noting from Theorem~\ref{Thm:optimal} that evaluating the robustness of a quantum classifier \(\a\)—specifically, computing its robust accuracy and identifying adversarial examples—requires invoking Algorithm~\ref{Algorithm} on each quantum state in the given dataset. Since the cost of a single call to Algorithm~\ref{Algorithm} is \(O(|\c| \cdot N^{6.5})\), the overall complexity for a dataset \(T\) of size \(|T|\) is 
\[
O(|T| \cdot |\c| \cdot N^{6.5}),
\]
where \(|\c|\) is the number of classes and \(N = \dim(\mathcal{H})\) is the dimension of the input state space.

However, the certified robustness lower bound provided in Theorem~\ref{thm:lower} offers a way to significantly reduce this computational cost. This bound can be computed efficiently in \(O(|\c| \cdot N^5)\) time, which corresponds to performing \(|\c|\cdot N^2\) matrix multiplications for \(N \times N\) density matrices by noting that the number of Kraus operator in a super-operator $\e$ can be limited to be less than $N^2$.

A practical strategy for efficient robustness verification is thus as follows: we first use the certified lower bound to pre-screen the training dataset \(T\), identifying all potentially non-robust states and collecting them in a subset \(T' \subseteq T\). We then apply the exact robustness-checking Algorithm~\ref{Algorithm} only to the states in \(T'\), using a separate set \(R\) to record the discovered adversarial examples and their indices.

This two-stage approach reduces the overall complexity to
\[
O(|T'| \cdot |\c| \cdot N^{6.5}),
\]
where typically \(|T'| \ll |T|\), as empirically confirmed by our simulation results in~\cite{guan2021robustness}. This demonstrates the practical effectiveness of using the robust lower bound for accelerating formal verification of robustness in quantum classifiers.

\subsection{Approximate Robustness Verification}

Finally, we introduce an efficient algorithm (Algorithm~\ref{algo:underapproximation}) for under-approximating the robust accuracy computed by Algorithm~\ref{algo:robustcheck}. Specifically, Algorithm~\ref{algo:underapproximation} serves as a lightweight subroutine that avoids calling an SDP solver whenever a potential non-robust state can be directly detected using the robustness lower bound established in Theorem~\ref{thm:lower}.

\begin{algorithm}[H]
\caption{UnderRobustAccuracy($\a,\epsilon,T$)}
\label{algo:underapproximation}
    \begin{algorithmic}[1]
    \Require  $\a=(\e,\{M_c\}_{c\in\c})$ is a well-trained quantum classifier, $\epsilon < 1$ is a real number, $T=\{(\rho_i,l_{i})\}$ is a dataset indicating $\rho_i$ be correctly classified into class $l_i$.
    \Ensure A  under-approximation of robust accuracy URA
    \State $r=0$.\Comment{Record the number of potential non-robust states}
    \ForAll{$(\rho_i,l_i)\in T$}
    \State Let \( p_c= \mathrm{Tr}[M_c \mathcal{E}(\rho_i)] \) be the measurement probability of observing outcome \( c \) for input \( \rho \).
    \State Obtain the lower bound $\varepsilon_{\mathrm{RLB}}(\rho_i)= \min_{c \ne c^*} \frac{1}{2} \left( \sqrt{p_{c^*}} - \sqrt{p_c} \right)^2$\Comment{Applying the robust bound in Theorem~\ref{thm:lower}}
    \If {$\epsilon>\varepsilon_{\mathrm{RLB}}(\rho_i)$} 
    \State $r=r+1$
    \EndIf
    \EndFor
    \State \Return URA$=1-\frac{r}{|T|}$.
    \end{algorithmic}  
\end{algorithm}

We have empirically compared Algorithm~\ref{algo:robustcheck} and Algorithm~\ref{algo:underapproximation} in terms of verification time and output accuracy across various quantum classifiers simulated on classical hardware, as reported in~\cite{guan2021robustness}. The results demonstrate that Algorithm~\ref{algo:underapproximation} exhibits favorable scalability and achieves significantly faster runtimes than Algorithm~\ref{algo:robustcheck}, while still providing accurate approximations of robust accuracy. These findings confirm the practical tightness of the proposed robustness lower bound.

\subsection{Verification Algorithm Comparison}

To conclude this section, we summarize the robustness verification algorithms based on different robustness bounds in Table~\ref{table:distinguishability}. These algorithms vary in computational complexity and verification guarantees. In practice, the condition $|T'| \ll |T|$ often holds, confirming the utility of robustness lower bounds in efficiently identifying non-robust quantum states.

\begin{table}[!htbp]
\centering
\scalebox{1}{
\begin{tabular}{|c|p{95pt}<{\centering}|p{95pt}<{\centering}|p{80pt}<{\centering}|}
\hline
\multicolumn{4}{|c|}{Robustness Verification Algorithms} \\ 
\hline
& Robustness Lower Bound & Robustness Optimal Bound & Mixed Strategy \\
\hline
Method & Matrix Multiplication (MM) & Semidefinite Programming (SDP) & MM \& SDP \\
\hline
Complexity & $O(|T| \cdot |\mathcal{C}| \cdot N^5)$ & $O(|T| \cdot |\mathcal{C}| \cdot N^{6.5})$ & $O(|T'| \cdot |\mathcal{C}| \cdot N^{6.5})$ \\
\hline
Robust Accuracy & Under-approximate & Exact & Exact \\
\hline
\end{tabular}
}
\caption{Summary of robustness verification algorithms based on different bounds.}
\label{table:distinguishability}
\end{table}

\vspace{1em}
Thanks to the linearity of quantum learning models—an outcome of the foundational postulates of quantum mechanics—robustness verification for quantum classifiers can often be performed efficiently, with polynomial-time complexity in the dimension of the input state. This is in stark contrast to classical machine learning models, such as deep neural networks (DNNs), which are highly nonlinear and non-convex. In such classical settings, even verifying simple properties can be NP-complete~\cite{katz2017reluplex}.

However, this computational advantage in the quantum setting diminishes when verification is restricted to \emph{pure states}. Unlike the convex set of mixed states, the set of pure states is non-convex. Consequently, the optimization problem underlying the computation of the robustness radius is no longer a semidefinite program (SDP) but becomes a \emph{quadratically constrained quadratic program} (QCQP)—an optimization class where both the objective and constraints are quadratic~\cite{guan2021robustness}. Solving QCQPs is NP-hard in general.

To address this challenge, Algorithm~\ref{Algorithm} can be adapted to pure-state robustness verification by replacing the SDP solver in Line~\ref{Algorithm:line:SDP} with a QCQP solver. The resulting variant can be integrated into Algorithm~\ref{algo:robustcheck} to compute robust accuracy and identify adversarial examples constrained to pure states. This adaptation has been evaluated in case studies such as MNIST classification, where input samples are encoded as pure quantum states~\cite{guan2021robustness}.

\section{VeriQR: A Tool for Robustness Verification}
\label{sec:veriqr}
To translate the robustness verification algorithms (Algorithms~\ref{algo:robustcheck} and~\ref{algo:underapproximation}) developed in the previous section into practical workflows, we introduce \textsc{VeriQR}~\cite{lin2024obustness}, the first dedicated software tool for certifying adversarial robustness in quantum machine learning. \textsc{VeriQR} bridges formal verification, adversarial evaluation, and robustness enhancement, providing a flexible and efficient platform for real-world quantum applications.

\textsc{VeriQR} is available at \url{https://github.com/Veri-Q/VeriQR}.

\begin{figure}[ht]
    \centering
    \includegraphics[width=1\linewidth]{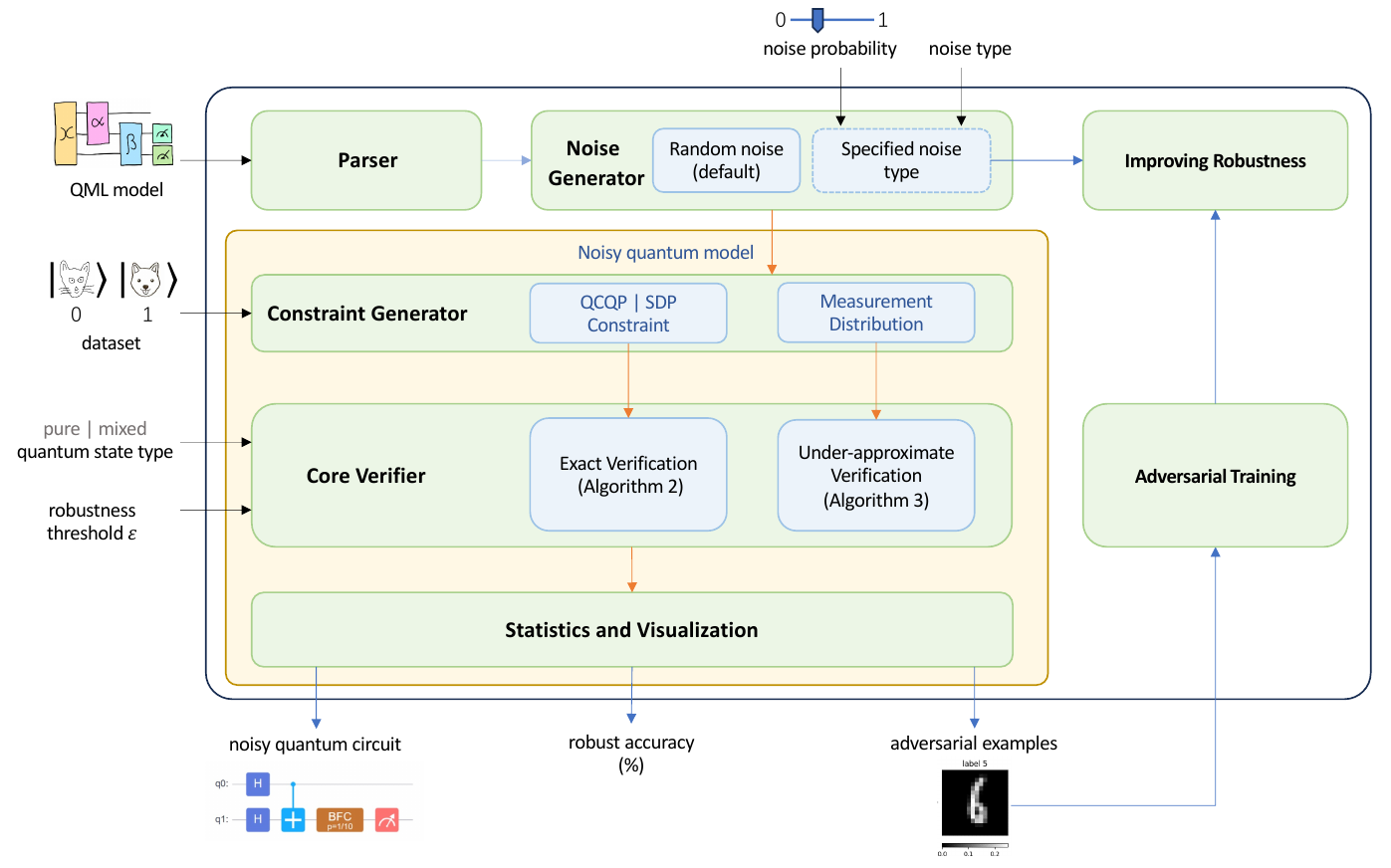}
    \caption{System architecture of \textsc{VeriQR}.}
    \label{fig:VeriQR}
\end{figure}

As illustrated in Fig.~\ref{fig:VeriQR}, the core capabilities of \textsc{VeriQR} include:
\begin{enumerate}
    \item \textit{Tool Integration:} A built-in parser compatible with OpenQASM, supporting frontends such as Qiskit, Cirq, and MindSpore Quantum;
    \item \textit{Noisy Simulation:} Injection of various noise models, including random noise, depolarizing, bit-flip, phase-flip, and user-defined noise;
    \item \textit{Robustness Verification:} Formal verification of adversarial robustness for quantum classifiers based on certified bounds;
    \item \textit{GUI and Visualization:} Interactive interface for diagnostics and visualizing robustness landscapes of quantum learning models;
    \item \textit{Robustness Improvement:} Enhancement of classifier robustness via adversarial training using counterexamples identified through formal verification or adding specified quantum noise.
\end{enumerate}
In the following, we detail these features.

\subsection{Setups and Inputs}\label{sec:overview}

\textsc{VeriQR} is a graphical user interface (GUI) tool implemented in C++, leveraging the widely adopted Qt framework for GUI development~\cite{blanchette2006c++}.

As shown in the upper left corner of Fig.~\ref{fig:VeriQR}, users begin by importing a QML model along with a dataset of quantum states and corresponding ground truth labels, drawn from either the training or testing phase. \textsc{VeriQR} supports models in the following formats, each representing a quantum circuit with a final measurement stage:

\begin{itemize}
\item[1.] \textit{NumPy data file (.npz format):} This format packages the quantum circuit, measurement operators, and dataset into a single file. It is particularly suitable for users familiar with classical machine learning and formal methods but less experienced in quantum computing. \textsc{VeriQR} includes several example .npz models, enabling new users to quickly begin robustness verification.

\item[2.] \textit{OpenQASM 2.0 file (.qasm format):} OpenQASM is a widely accepted standard for describing quantum circuits, originally introduced by IBM~\cite{cross2017open}. QML models trained using various quantum platforms can be converted into this format for unified robustness analysis within \textsc{VeriQR}.
\end{itemize}

Once a model is loaded, users can configure verification parameters via the control panel, as depicted in the left and top portions of Fig.~\ref{fig:VeriQR}. The configurable options include:

\begin{enumerate}
\item \textbf{Noise Model.} Select among several quantum noise types, including \textit{depolarizing}, \textit{phase-flip}, and \textit{bit-flip} noise~\cite{nielsen2010quantum}. Random noise is enabled by default. Additionally, users may define custom noise models or apply combinations of different types.
\item \textbf{Quantum State Type.} Specify whether the input quantum states are \textit{mixed} or \textit{pure}. By default, \textsc{VeriQR} assumes mixed states, as they better represent realistic NISQ system states.

\item \textbf{Robustness Threshold.} Set the perturbation tolerance parameter $\epsilon$ for robustness verification. This determines the minimum fidelity distance required to guarantee robustness certification.
\end{enumerate}

\subsection{Verifying Robustness}
\begin{figure}[h]
    \vspace{-0.4cm}
    \centering
    \includegraphics[width=\linewidth]{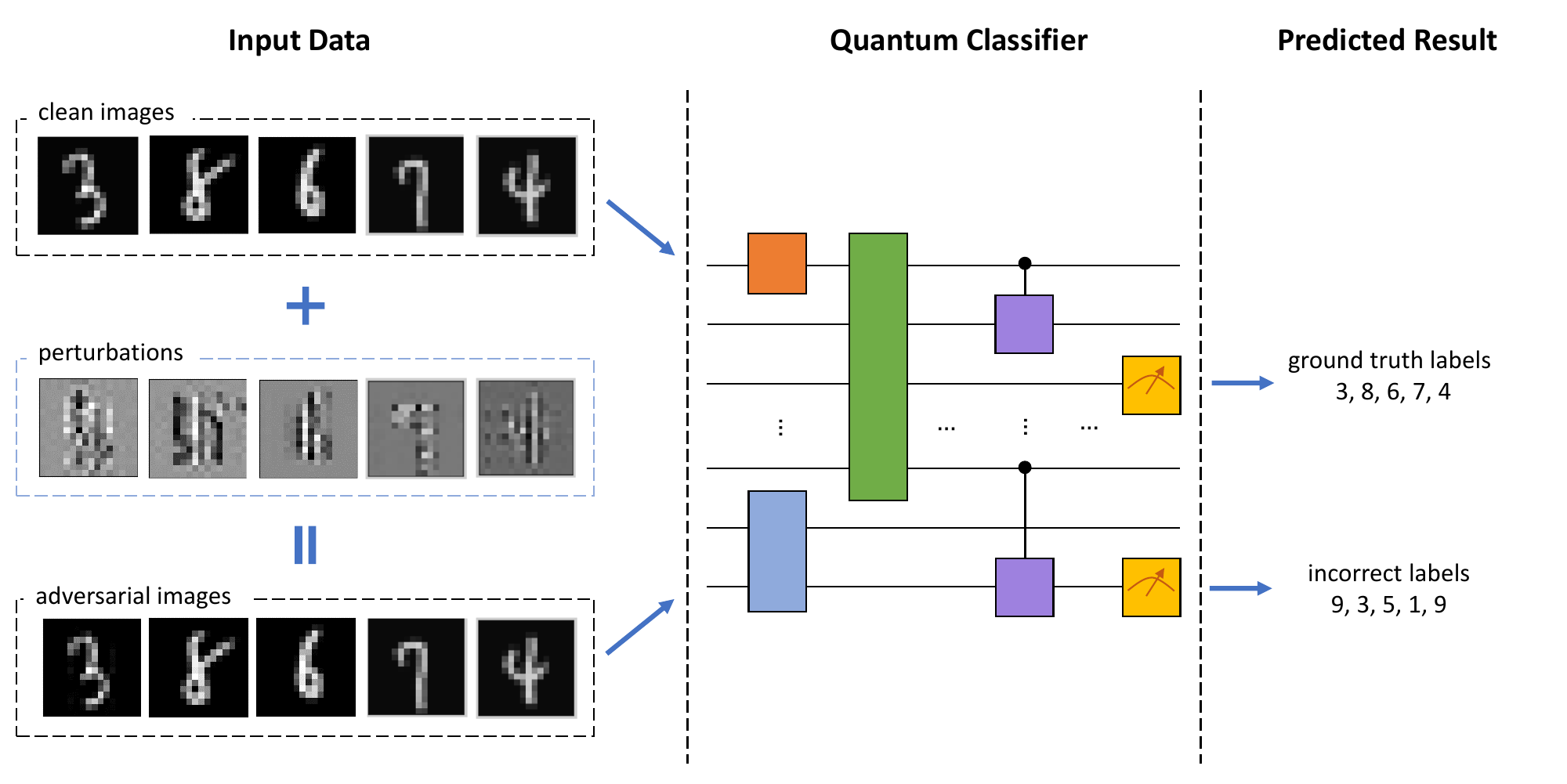}
    \caption{The adversarial examples and the corresponding adversarial perturbations found by \textsc{VeriQR} in MNIST handwritten digit classification. }
    \label{fig:adv_example}
    \vspace{-0.8cm}
\end{figure}

As shown in the center of Fig.~\ref{fig:VeriQR}, the robustness verification workflow in \textsc{VeriQR} consists of five key modules:

\begin{itemize}
\item[1)] \textbf{Parser.} This module reads the quantum classifier file and constructs the corresponding quantum circuit object for downstream analysis.
\item[2)] \textbf{Noise Generator.} Taking a quantum circuit as input, this module supports two noise injection modes:
\begin{itemize}
    \item In the default mode, random noise is applied to each qubit at random positions within the circuit, with randomly sampled noise probabilities, simulating the effects of real-world quantum hardware noise.
    \item Alternatively, users can specify noise types (e.g., depolarizing, bit-flip, phase-flip, or custom Kraus operators) and inject them at the end of the circuit, a commonly adopted assumption. This functionality also enables robustness enhancement, as discussed in~\cite{lin2024obustness}.
\end{itemize}

\item[3)] \textbf{Constraint Generator.} This module generates the necessary mathematical constraints from the (possibly noisy) quantum circuit and dataset, preparing them for the verification engine.

\item[4)] \textbf{Core Verifier.} Given the generated constraints, a perturbation threshold $\varepsilon$, and the type of quantum state (mixed or pure), this module selects an appropriate solver: an SDP solver for mixed states or a QCQP solver for pure states~\cite{guan2021robustness}. It executes both exact and under-approximate verification procedures via Algorithms~\ref{algo:robustcheck} and~\ref{algo:underapproximation}, respectively, to certify $\varepsilon$-robustness.

\item[5)] \textbf{Statistics and Visualization.} This module visualizes verification results in the GUI. It reports the robust accuracy of the classifier and stores detected adversarial examples in a NumPy file for further analysis or adversarial training. Additionally, it displays the quantum circuit diagrams before and after noise injection. For MNIST digit classification tasks, \textsc{VeriQR} also presents images of adversarial examples corresponding to digits selected by the user, as illustrated in Fig.~\ref{fig:adv_example}.
\end{itemize}

\subsection{Improving Robustness}\label{sec:improving_robustness}

\textsc{VeriQR} supports two complementary strategies to enhance the adversarial robustness of QML models: adversarial training and the deliberate injection of specific quantum noise.

{\textbf{Adversarial Training.}} \textsc{VeriQR} extends classical adversarial training techniques to the quantum domain. When the $\epsilon$-robustness of a state $\rho$ with true label $l$ fails, the embedded verification algorithms automatically generate an adversarial example $\sigma$. The pair $(\sigma, l)$ can then be added to the training dataset, allowing the QML model to be retrained with adversarial samples. This iterative procedure effectively improves the model’s robustness against future adversarial perturbations.

{\textbf{Injection of Specific Noise.}}
Prior studies~\cite{du2021quantum,guan2022verifying,huang2023certified} have shown that carefully introducing quantum noise at selected points within a circuit can enhance robustness, potentially outperforming random noise. \textsc{VeriQR} provides users with the ability to inject either standard or user-defined noise models—using Kraus operators—at specific positions in the quantum circuit. This functionality enables systematic exploration and optimization of noise-assisted robustness strategies for QML models.

\section{Experimental Benchmark on Superconducting Hardware}\label{sec:benchmark}

To evaluate the practical hardware effectiveness of our robustness verification framework, the first experimental benchmark of adversarial robustness in QML was conducted on real superconducting quantum hardware~\cite{zhang2025experimental}. This section outlines the experimental setup, summarizes the benchmark results, and highlights insights obtained from physical validation.

\begin{figure*}[ht]
	\centering
	\includegraphics[width=\textwidth]{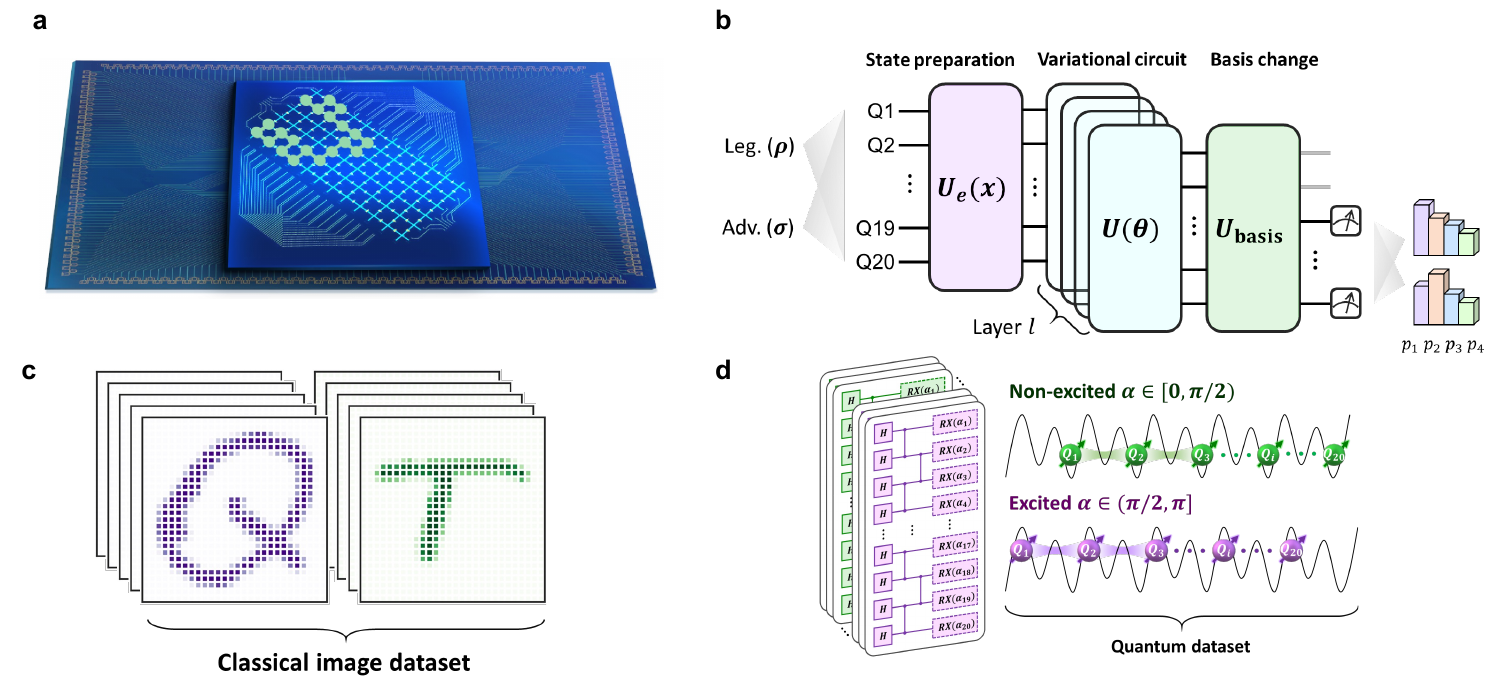}
	\caption{\textbf{Experimental schematic for QNN robustness evaluation.} 
	\textbf{a}, Schematic diagram of the superconducting quantum processor, comprising 72 qubits and 126 couplers arranged in a 2D lattice. The 20 qubits selected for the experiment are highlighted in green. 
	\textbf{b}, Architecture of the quantum neural network (QNN) classifier, including the state preparation circuit, an $l$-layer variational circuit, and pre-measurement basis transformation gates. 
	\textbf{c}, Sample visualization of handwritten letters ``Q'' and ``T'' from the EMNIST dataset, used for the classical image classification task. 
	\textbf{d}, Quantum circuit used to generate the LCEI dataset, showing the application of an $R_x(\alpha)$ rotation following the linear cluster state. States are labeled as ``excited'' or ``non-excited'' based on the rotation angle $\alpha$.}
	\label{fig:setup}
\end{figure*}

The formal framework was validated by executing QML classifiers on a 20-qubit superconducting processor~\cite{zhang2025experimental}. As shown in Fig.~\ref{fig:setup} \textbf{a}, the processor comprises 72 qubits and 126 couplers arranged in a 2D lattice; 20 qubits with high fidelity, highlighted in green, were selected for the experiments.

Two types of classification tasks were considered—classical and quantum—both implemented using trained quantum neural network (QNN) classifiers. As depicted in Fig.~\ref{fig:setup} \textbf{b}, each QNN architecture consists of three components: a state preparation circuit that encodes data into a quantum input state, a variational circuit that performs the learning task, and a basis transformation followed by quantum measurement to extract output predictions.

\begin{enumerate}
\item \textbf{Classical Task:} The classification of handwritten characters Q'' and T’’ was performed using the \emph{EMNIST dataset}\cite{cohen2017emnist}, as illustrated in Fig.\ref{fig:setup} \textbf{c}.
\item \textbf{Quantum Task:} The task of \emph{Linear Cluster State Excitation Identification (LCEI)}~\cite{guan2021robustness,broughton2020tensorflow} involved distinguishing between excited and non-excited $20$-qubit cluster states. These states were generated using a synthetic dataset, with excitation determined by the rotation angle $\alpha$ of an $R_x(\alpha)$ operation applied at the end of the cluster state preparation circuit (Fig.~\ref{fig:setup} \textbf{d}).
\end{enumerate}

For both tasks, classification was based on the expectation value of the output qubit with respect to the Pauli-Z operator, $\sigma_z$~\cite{nielsen2010quantum}. The resulting expectation value was converted into a probability via the transformation $p = (\langle \sigma_z \rangle + 1)/2$, with a threshold of $p = 0.5$ used to define the decision boundary.


For each classification task, adversarial robustness benchmarking is conducted on a well-trained binary quantum classifier using 10 randomly selected input quantum states (5 from each class). The experimental procedure involves the following steps:

\begin{enumerate}
\item \textbf{Measurement:} For each selected input quantum state $\rho$, execute the quantum classifier circuit on the quantum hardware to obtain the binary measurement outcome probabilities ${p_1, p_2 = 1 - p_1}$. Without loss of generality, it is assumed that $p_1 \geq p_2$ to facilitate subsequent analysis.
\item \textbf{Robustness Lower Bound:} Compute the robustness lower bound $\varepsilon_{\mathrm{RLB}}(\rho)$ based on the measured probabilities using Theorem~\ref{thm:lower}.

\item \textbf{Robustness Upper Bound:} Apply the Mask FGSM attack (as described in Section~\ref{sec:upper_bound}) on the quantum hardware to estimate the robustness upper bound $\varepsilon_{\mathrm{RUB}}(\rho)$ and identify the corresponding adversarial example $\sigma_{\rm adv}$.

\item \textbf{Adversarial Training:} Retrain the quantum neural network classifier by incorporating the adversarial examples of all 10 quantum states into the training set, using their correct labels, in order to enhance the model's robustness.

\item \textbf{Adversarial Analysis:} Compare the robustness bounds before and after adversarial training to evaluate the tightness and stability of the certification framework.
\end{enumerate}

   \begin{figure*}[!th]
	\centering
	\includegraphics[width=1\textwidth]{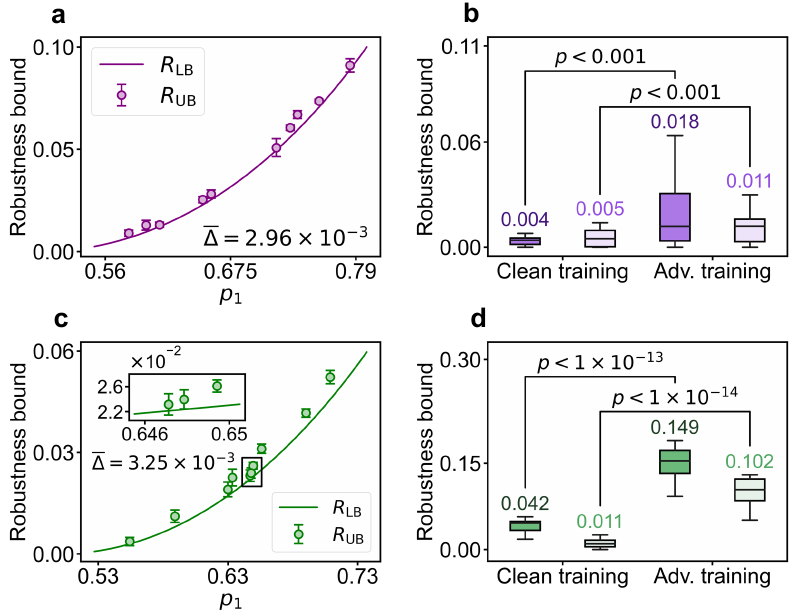}
	\caption{\textbf{Robustness bound verification experiments.} $\textbf{a}$, $\textbf{c}$, Comparison of the experimental upper bound $R_{\rm{UB}}$ (from $10$ randomly selected samples, $5$ per class) versus theoretical $R_{\rm{LB}}$. Error bars indicate the root mean square error from fitting $D(\rho,\sigma)$. $\overline{\Delta}$ denotes the average gap between $R_{\rm{UB}}$ and $R_{\rm{LB}}$ of the $10$ samples. $\textbf{b}$, $\textbf{d}$, The robustness bounds for critical samples under clean and adversarial training. Adversarial training significantly increased the average robustness lower bound, with dark and light colors denoting distinct classes. $p$-value $\leq$ $0.001$, indicating statistically significant differences. Panels $\textbf{a}$ and  $\textbf{b}$ correspond to EMNIST dataset, while $\textbf{c}$ and $\textbf{d}$ to LCEI.}
	\label{fig:results}  
\end{figure*}
As shown in Fig.~\ref{fig:results}, the experimental results validate two key aspects of the robustness verification framework:

\begin{enumerate}
\item \textbf{Tightness of Robustness Bounds:} Fig.\ref{fig:results} \textbf{a} and \textbf{c} compare the experimentally measured robustness upper bounds $\varepsilon_{\rm RUB}$ (denoted as $R_{\rm UB}$) against the certified robustness lower bounds $\varepsilon_{\rm RLB}$ (denoted as $R_{\rm LB}$), computed from five randomly selected samples per class. In both EMNIST and LCEI tasks, the upper bounds consistently exceed the corresponding lower bounds, with average gaps of $2.96 \times 10^{-3}$ and $3.25 \times 10^{-3}$, respectively. These results demonstrate the near-optimality of the Mask FGSM attack strategy and affirm the tightness of the certified lower bounds derived in Theorem~\ref{thm:lower}.
\item \textbf{Improvement through Adversarial Training:} Critical samples, which are defined as the $20\%$ of instances with the lowest robustness under clean training, were used to evaluate the effect of adversarial training. As depicted in Fig.~\ref{fig:results}~\textbf{b} and \textbf{d}, adversarial training significantly increased the mean certified robustness lower bound for these critical samples, by a factor of $4.22$ in EMNIST and $4.74$ in LCEI. This improvement underscores the utility of adversarial training in enhancing the robustness of QMLs and mitigating misclassification under adversarial perturbations.
\end{enumerate}

\section{Summary and Outlook}
\label{sec:outlook}

This review has presented a comprehensive overview of our recently developed formal verification framework for certifying the adversarial robustness of quantum machine learning (QML) classifiers. The framework unifies rigorous theoretical foundations, algorithmic advances, toolchain development, and experimental validation into an end-to-end pipeline for trustworthy QML on quantum devices.

At the theoretical level, the framework introduces two certified robustness bounds for quantifying the adversarial robustness radius of a QML classifier at a given quantum input state:
\begin{itemize}
    \item \textbf{Optimal Robustness Bound.} This bound exactly quantifies the adversarial robustness radius and can be computed via semidefinite programming (SDP) when the input state is mixed, and via quadratically constrained quadratic programming (QCQP) for pure states.
    \item \textbf{Robustness Lower Bound.} This efficiently computable bound, derived from measurement statistics, certifies a region around a given input state where the classifier's output remains unchanged. It is especially valuable for identifying potentially non-robust inputs with low computational overhead and is well suited for hardware-level evaluation, as it does not require access to the internal evolution of the quantum classifier.
\end{itemize}

These certified bounds enable both exact and under-approximate verification. Furthermore, they are connected through the \textbf{robustness upper bound}, an empirical robustness estimate derived from adversarial attacks (e.g., the Mask FGSM method). This empirical bound serves as a benchmark for evaluating the tightness of the certified lower bound in practical scenarios.

To operationalize the theoretical results, the software tool \textsc{VeriQR} has been developed, serving as the first dedicated platform for robustness verification of QML models. \textsc{VeriQR} supports exact and approximate verification, simulates various noise models, enables adversarial training, and integrates with major quantum programming frameworks (e.g., Qiskit, Cirq). A graphical user interface (GUI) supports intuitive robustness diagnostics and visualization.

Notably, this framework has been experimentally validated on a 20-qubit superconducting quantum processor. These experiments constitute the first hardware-based benchmark of adversarial robustness in QML. The results show that:
\begin{itemize}
    \item The certified robustness lower bounds closely match empirical upper bounds across classification tasks.
    \item Adversarial training substantially improves the robustness guarantees of quantum classifiers, particularly for inputs most vulnerable to perturbation.
\end{itemize}

With these results, our framework marks a promising step toward trustworthy QML on noisy intermediate-scale quantum (NISQ) hardware and opens promising directions for scalable and hardware-integrated robustness certification. In the following, we outline several key avenues for future research and development.

\textbf{1. Hardware-integrated robustness verification.}
While the current implementation of \textsc{VeriQR} operates within classical simulation environments, a critical next step is to extend the framework for execution directly on quantum hardware. This would enable robustness certification under real-world conditions, accounting for hardware-specific noise characteristics, calibration drift, and temporal fluctuations. Such integration would bridge the practical gap between formal robustness guarantees and physical device behavior, enabling more faithful certification of QML models in practical deployment scenarios.

\textbf{2. Certification of dynamically trained quantum classifiers.}
Most existing robustness verification approaches, including those presented in this review, assume a fixed and pre-trained quantum classifier. However, many real-world applications require models that learn from data in real time or adapt to adversarial perturbations. A promising direction is to develop verification methods that support dynamic or online QML training. In this setting, the quantum model evolves continuously, and robustness certification must be integrated into the training process to ensure that robustness guarantees are preserved or improved over time. This requires the design of adaptive certification techniques that can operate in tandem with quantum training loops, like their classical counterparts~\cite{meng2022adversarial}.

\textbf{3. Beyond robustness: certifying fairness, interpretability, and privacy.}
While this framework focuses on adversarial robustness, many other aspects of trustworthiness are equally important in quantum AI systems. Future research may extend the verification methodology to properties such as fairness across quantum finance, interpretability of quantum decision-making, and differential privacy under quantum data access. Initial efforts in these directions have already begun~\cite{guan2022verifying,guan2023detecting,guan2024optimal}, leveraging the robustness-based verification framework. A unified formal foundation for verifying such multifaceted trust properties would significantly strengthen the overall reliability of QML.

\textbf{4. Cross-platform robustness benchmarking.}
The experimental results in this work are based on superconducting quantum hardware. Expanding the robustness benchmarking to other quantum platforms, such as trapped ions, neutral atoms, or photonic systems, would offer deeper insights into the relationship between hardware noise characteristics and model robustness. Establishing standard robustness benchmarks across different architectures will be essential for evaluating and comparing the trustworthiness of QML implementations on heterogeneous quantum systems.

\vspace{0.5em}
These directions outline a broader agenda for building verifiable, trustworthy, and scalable quantum AI systems. As quantum computing advances toward practical and widespread use, formal verification tools like those presented in this framework may play an important role. By embedding trust guarantees into the lifecycle of quantum machine learning from model design and training to deployment and adaptation, future research can ensure that quantum AI not only performs well, but also behaves reliably and predictably in complex and uncertain environments. It is hoped that this review will foster broader engagement from the formal methods and quantum AI communities toward this critical goal.

\bibliographystyle{unsrt}
\bibliography{main}
\end{document}